\documentclass[11pt]{article}

\usepackage[preprint]{acl}

\usepackage{times}
\usepackage{latexsym}
\usepackage{amssymb}
\usepackage{booktabs}
\usepackage{multirow}
\usepackage{tabularx}
\usepackage[bb=boondox,bbscaled=.95,cal=boondoxo]{mathalfa} 
\usepackage{amsmath} 
\usepackage{enumitem}
\usepackage[T1]{fontenc}

\usepackage[utf8]{inputenc}

\usepackage{microtype}

\usepackage{inconsolata}

\usepackage{graphicx}
\usepackage{subfig}
\title{Towards Multi-Agent-Simulation-Based Community Note Evaluation}

\author{
  \textbf{Changxi Wen}\textsuperscript{1}\thanks{~~Equal contribution.},
  \textbf{Shuning Zhang}\textsuperscript{1}\footnotemark[1],
  \textbf{Bohao Chu}\textsuperscript{2},
  \textbf{Yuwei Chuai}\textsuperscript{3}\thanks{~~Corresponding authors.},
\\
  \textbf{Hui Wang}\textsuperscript{2},
  \textbf{Dai Shi}\textsuperscript{4},
  \textbf{Xin Yi}\textsuperscript{1}\footnotemark[2],
  \textbf{Hewu Li}\textsuperscript{1}
\\
\\
  \textsuperscript{1}Tsinghua University, Beijing, China \\
  \textsuperscript{2}University of Duisburg-Essen, Duisburg, Germany \\
  \textsuperscript{3}University of Luxembourg, Luxembourg \\
  \textsuperscript{4}Tongji University, Shanghai, China \\
}

\begin{document}
\maketitle
\begin{abstract}
Community-based fact-checking that relies on cross-consensus is expanding rapidly on social media platforms. However, the delay and low-ratio of cross-consensus community fact-checks rated by human contributors remains a significant challenge. To address this, we first created ComRate, a large-scale dataset comprising 2.5 million community notes and over 209 million ratings sourced from $\mathbb{X}$. We then propose MultiCom, a persona-guided multi-agent rating framework for community note evaluation. MultiCom simulates diverse rater population by clustering contributors in a matrix-factorized rater space and prompting persona agents to generate structured assessments based on the official community notes rating schema. These agents output structured and explainable judgments, such as confidence, agreement signals and reasons. An out-of-fold calibrated aggregation algorithm combines features such as raw votes and diagnostic reason signals for reliable prediction. Extensive evaluations demonstrate that MultiCom outperforms alternative methods, achieving an average accuracy of 84.7\% (balanced accuracy 68.3\%, macro-F1 60.1\%) on the evaluation set.

\end{abstract}

\section{Introduction}

Tackling misinformation and disinformation remains a critical priority for social platforms. While early initiatives relied on professional fact-checkers~\cite{micallef2022true} or automated fact-checking systems~\cite{guo2022survey}, these approaches often face high costs and limited scalability. Crowdsourced fact-checking has emerged as a scalable alternative, leveraging collective efforts to author ``community notes'' -- short, evidence-based contexts designed to debunk misleading posts~\cite{prollochs2022community} and curb the spread of misinformation~\cite{chuai2024did,chuai2026request}. Such programs have been operational on platforms like $\mathbb{X}$ for over five years, spanning from early 2021 to 2026.

However, the debunking community notes still needs human raters to determine their actual helpfulness~\cite{chuai2026community,prollochs2022community}. In fact, this is not unique to such crowdsourced fact-checking systems. Even professional fact-checkers engage in cross-checking when doing fact-check work. These work are shown to improve fact-checks' comprehensiveness~\cite{warren2025show,micallef2022true}. In the era of Generative AI, where automated fact-checking are increasingly proposed and adopted~\cite{nakov2021automated,guo2022survey}, how to evaluate the generated fact-checking materials is important, especially from a user-centric perspective.

Existing literature predominantly focuses on automated note generation~\cite{de2025supernotes,zhang2025commenotes}, scaffolding note-writing workflows~\cite{xing2026communitynotes}, or general automated fact-checking architectures~\cite{nakov2021automated,guo2022survey}. Crucially, the sparse research addressing note evaluation~\cite{xing2026communitynotes} fails to account for complex evaluation dynamics, such as intermediate rating statuses.

Towards these challenges, we first construct ComRate, a large-scale real-world dataset comprising 209,290,533 ratings towards 2,566,644 community notes, spanning 1,698,835 posts from Jan 2021 to Apr 2026. We then propose MultiCom, a persona-guided multi-agent rating algorithm designed for debunking note evaluation. MultiCom leverages matrix factorization to cluster agents that mimic heterogeneous human rater personas. These agents perform explainable reasoning across multiple nuanced quality dimensions, such as evidence strength and claim coverage, instead of providing simple binary labels. Finally, a selective aggregation agent employs cross-validation and dual-threshold decision rules to prioritize reliable outcomes and resolve ``needs more ratings'' cases.

Extensive evaluations on ComRate demonstrate that MultiCom outperforms alternative methods. It further  generalizes to unseen future notes, and notes with different reasons. Our contributions are three-fold:

\begin{itemize}[leftmargin=*,noitemsep,topsep=0pt]
    \item \textit{Method:} We introduce MultiCom, a multi-agent framework that uses persona-guided simulation and multi-dimensional reasoning for explainable note evaluation.
    \item \textit{Dataset:} We provide ComRate, the most comprehensive real-world rating dataset of community notes and human ratings.
    \item \textit{Empirical:} We demonstrate MultiCom's effectiveness, generalizability across time and models, and its ability to provide diagnostic feedback for improving fact-checking quality.
\end{itemize}

\section{ComRate}

We constructed the ComRate dataset using data from the $\mathbb{X}$ API and the platform's official open-source repository\footnote{\url{https://communitynotes.x.com/guide/en/under-the-hood/download-data}}. The resulting dataset comprised 209,290,533 ratings on 2,566,644 community notes attached to 1,698,835 posts. The data spans a five-year period from January 28, 2021 to April 5, 2026.

To provide insights into the fact-checking ecosystem, we conducted an analysis of the dataset (Fig~\ref{fig:analysis}, detailed methodology see Appendix~\ref{sec:details_dataset}). First, the temporal distribution highlights a rapid adoption and scaling of the program, with the volume of notes, posts, and ratings peaking prominently in 2024 (Fig~\ref{fig:analysis} (a)). Second, Fig~\ref{fig:analysis} (b) provides a heatmap of standardized behavioral profiles across distinct rater clusters, with feature definitions and z-score normalization detailed in Appendix~\ref{sec:details_dataset}. This underscores the necessity of modeling diverse evaluator personas instead of assuming a uniform rater population. Third, we examined the distribution of misinformation categories and note-to-post ratios (Fig~\ref{fig:analysis} (c)). We found ``Factual error'' and ``Manipulated media'' are the most frequent categories, and the vast majority of posts have a single note. 

Additional analyses including language distribution and character-length statistics, are detailed in Appendix~\ref{sec:appendix_dataset_statistics}. 
Since the official Community Notes release does not provide complete post text for all notes, these additional full-dataset statistics are computed from note text and official note metadata.

Our task focuses on predicting whether or not a note is helpful. This aligns with the criteria of Community Note program on $\mathbb{X}$. We define helpful as \textbf{whether the note provides important context that helps a person recontextualize the original post}. For a note to be helpful, it should satisfy the following dimensions, as recommended by $\mathbb{X}$. It should be \textit{well-sourced} (relevant and high-quality citations), \textit{clear} (easily understandable language), \textit{comprehensive} (addressing all key claims), \textit{relevant} (providing crucial context), and \textit{neutral} (free from argumentative, speculative, or biased rhetoric). 

Our evaluation classify each note into one of three statuses, as in $\mathbb{X}$: \texttt{Helpful} (denoted as \texttt{H}), \texttt{Not Helpful} (\texttt{NH}), and \texttt{Needs More Ratings} (\texttt{NMR}). We denote \texttt{H}/\texttt{NH} as \textit{resolved status}, and retain \texttt{NMR} as some notes are contested and could not be simply classified into binary classes. 

\begin{figure}[!htbp]
    \centering

    \subfloat[Temporal growth of community notes records.]{
        \includegraphics[width=0.96\linewidth]{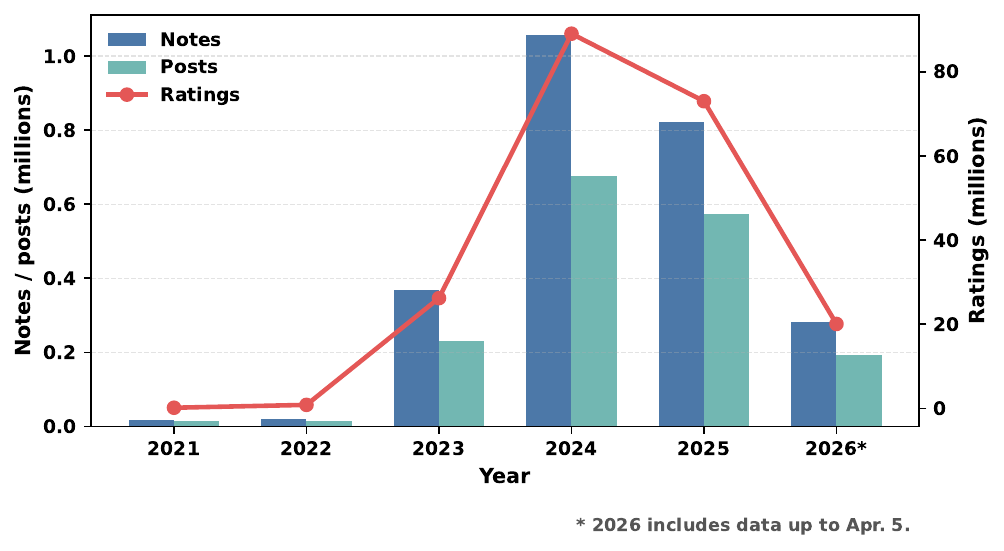}
        \label{fig:temporal_trend}
    }

    \subfloat[Rater-cluster behavioral profiles.]{
        \includegraphics[width=0.96\linewidth]{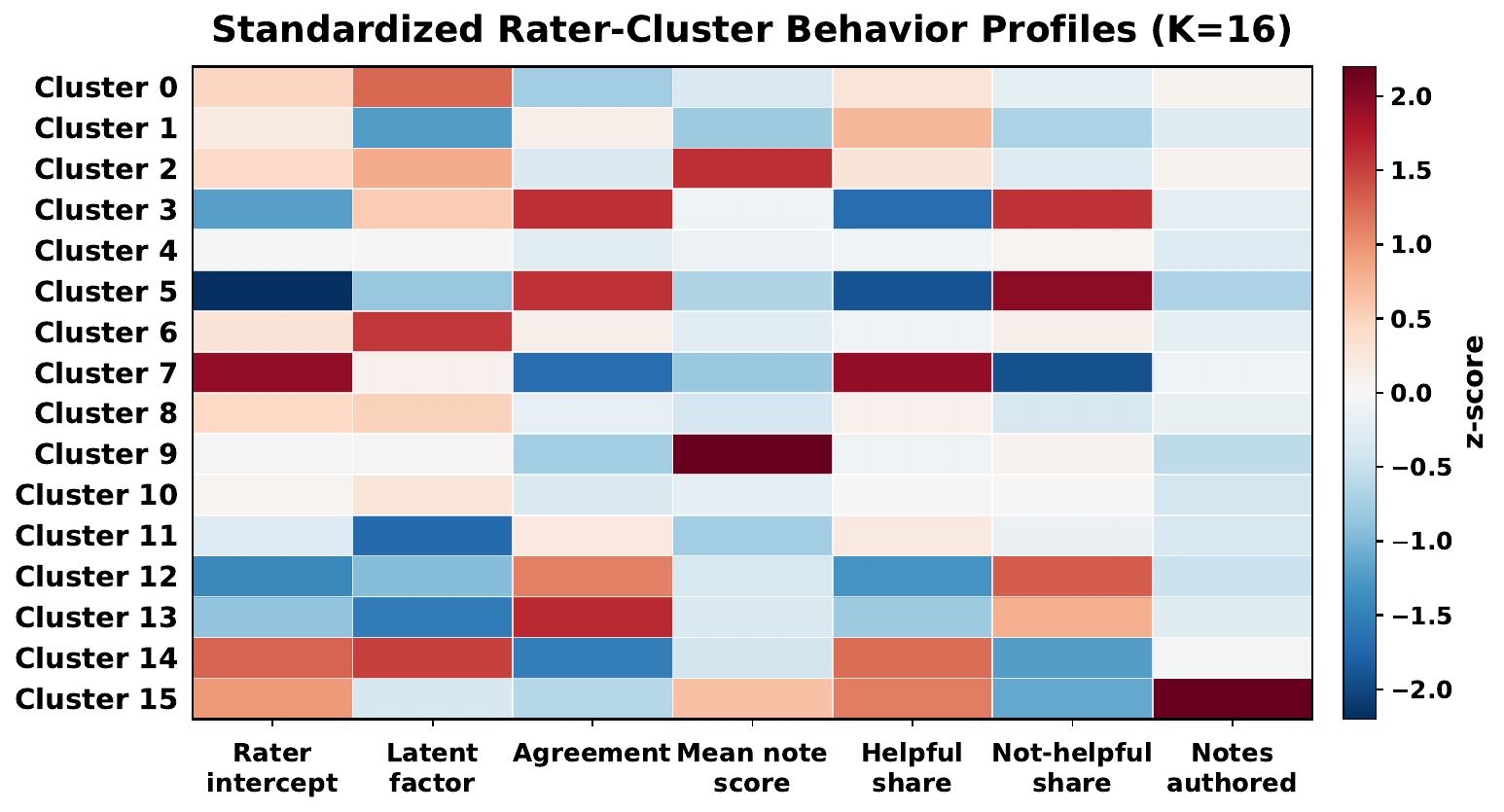}
        \label{fig:cluster_insight}
    }

    \subfloat[Distributions of note categories and notes per post.]{
        \includegraphics[width=0.96\linewidth]{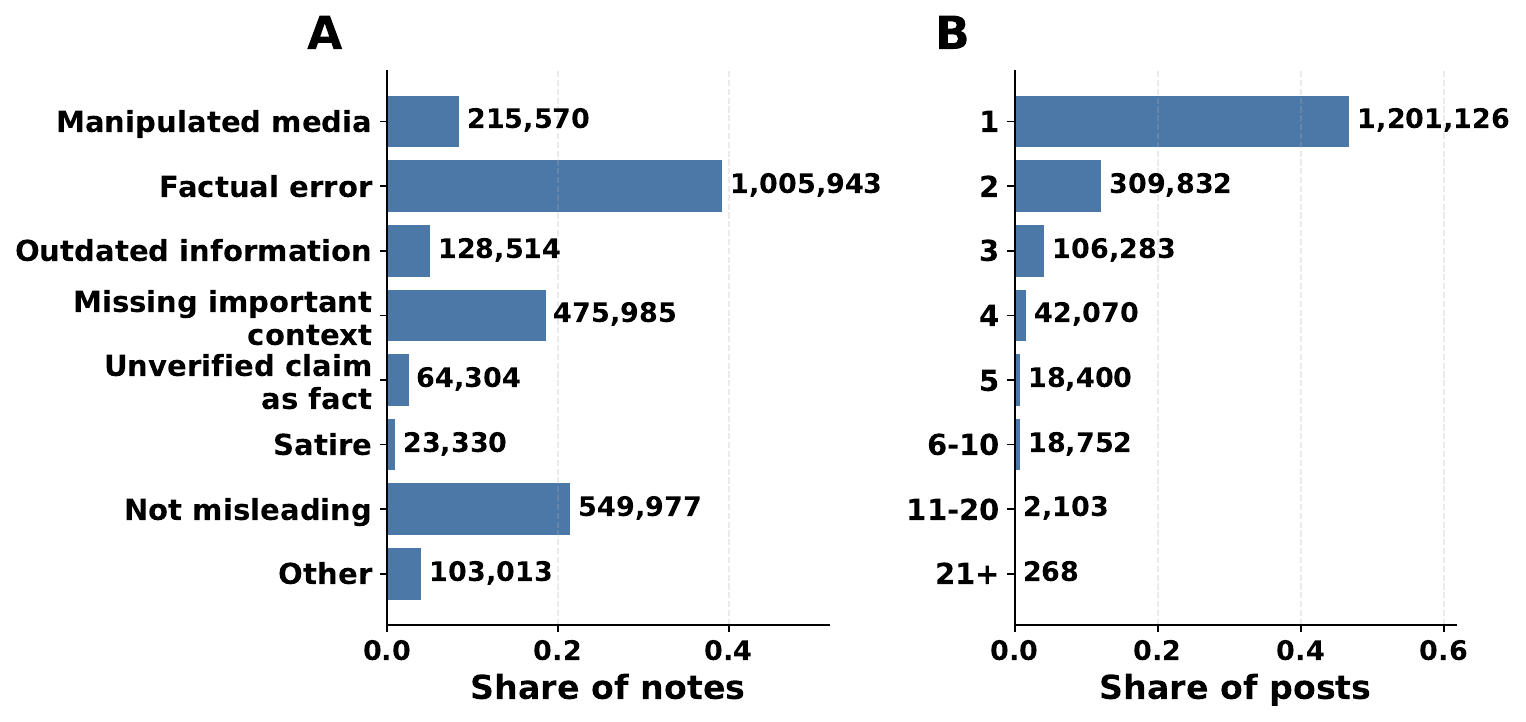}
        \label{fig:full_dataset_distributions}
    }

    \caption{
    Descriptive analysis of ComRate dataset.
    }
    \label{fig:analysis}
\end{figure}

\section{MultiCom}\label{sec:multicom}

\subsection{Algorithm Pipeline}

\begin{figure*}[!htbp]
    \centering 
    \includegraphics[width=\textwidth]{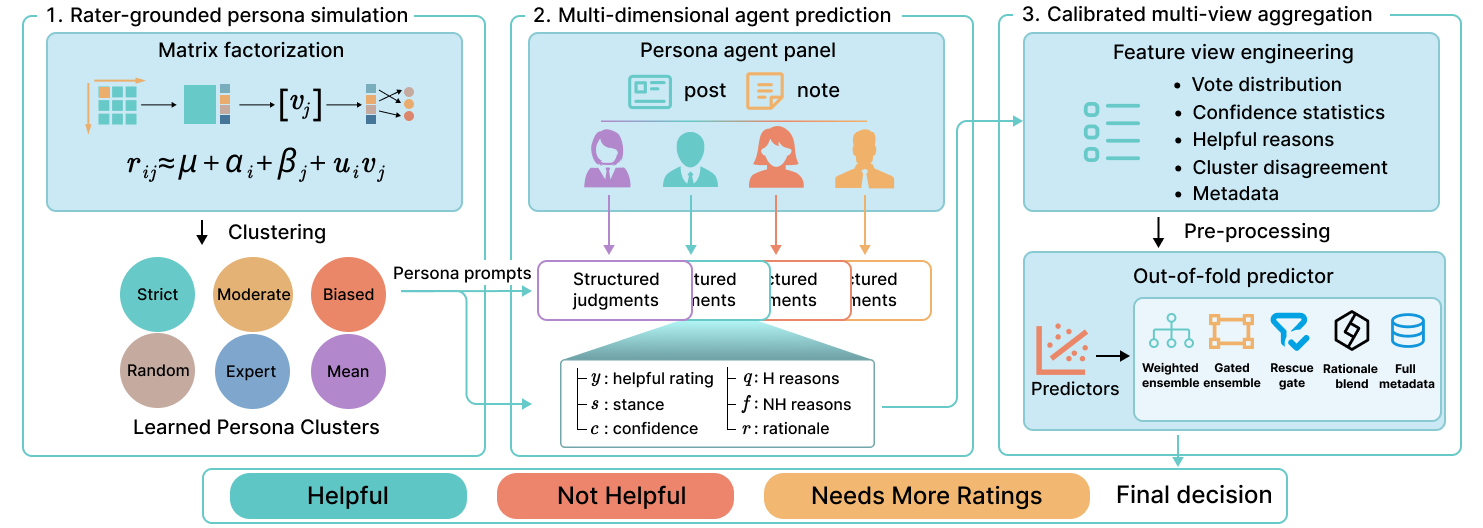}
    \caption{The algorithm flow of MultiCom.}
    \label{fig:pipeline}
\end{figure*}

We design the pipeline as a multi-agent system, emphasizing diversity, diagnostic judgment, and reliable final decisions, as shown in Figure~\ref{fig:pipeline}. we adopted a multi-agent simulation structure for rating, which constructs cluster-grounded agents from a matrix-factorization structure, so that the agents reflect heterogeneity during the rating process. For each agent, we ask it to produce multi-dimensional judgments instead of only a binary label, as helpfulness depends on multi-faceted aspects such as stance and evidence quality. Preserving these information could increase aggregation effectiveness and provide explainable reasoning. Finally, the aggregation module decide on cluster-level and agent-level features to reliably aggregate results. 

\subsection{Rater-Grounded Persona Simulation}

To analyze and learn raters' persona, we first learn a contributor space with biased rank-one matrix factorization,

\begin{equation}
    r_{ij} \approx \mu + \alpha_i + \beta_j + u_i v_j
\end{equation}

where $r_{ij}$ denotes the observed rating from contributor $i$ to note $j$, $\mu$ is the global intercept, $\alpha_i$ and $\beta_j$ are contributor- and note-specific biases, and $u_i$ and $v_j$ are one-dimensional latent factors. We then cluster contributors in this learned rater space to obtain distinct behavioral groups. 
For each cluster, we summarize its empirical rating profile, including the cluster size, mean historical helpfulness rating, tendency to rate helpful/not-helpful, agreement tendency, and reason-selection patterns when such reason annotations are available. 
Each cluster is then converted into a persona prompt that instructs the corresponding agent to evaluate notes according to this cluster-level rating behavior. 
In this way, agents simulate empirically observed rater groups with different  agreement tendencies, helpfulness priors, strictness levels, and sensitivities to note-quality dimensions (e.g., source quality, claim coverage).

\subsection{Multi-dimensional Agent Prediction}




Each agent outputs a structured judgment following the Community Notes program schema. 
For a post-note pair $(p,n)$, agent $a$ produces

\begin{equation}
z_a(p,n) =
(y_a, \mathbf{s}_a, c_a, \mathbf{q}_a, \mathbf{f}_a, r_a),
\end{equation}

where $y_a \in \{\texttt{helpful}, \texttt{somewhat helpful}, \texttt{not helpful}\}$ is the agent's overall helpfulness rating. 
The stance vector $\mathbf{s}_a$ contains the agent's agreement signals, including ``agree'' and ``disagree''. 
The confidence signal $c_a$ records how confident the agent is in its judgment. 
The quality vector $\mathbf{q}_a$ contains helpfulness reasons used in community notes' ratings: \texttt{Clear}, \texttt{GoodSources}, \texttt{AddressesClaim}, \texttt{ImportantContext}, and \texttt{UnbiasedLanguage}. 
These dimensions capture whether the note is clear, well-supported, directly addresses the claim, provides important context, and uses neutral language. 
The failure vector $\mathbf{f}_a$ contains not helpful reasons: \texttt{Incorrect}, \texttt{Sources\allowbreak Missing\allowbreak Or\allowbreak Unreliable}, \texttt{Missing\allowbreak Key\allowbreak Points}, \texttt{Hard\allowbreak To\allowbreak Understand}, \texttt{Argumentative\allowbreak Or\allowbreak Biased},\texttt{Irrelevant\allowbreak Sources}, \texttt{Opinion\allowbreak Speculation}, and \texttt{Note\allowbreak Not\allowbreak Needed}. 
These variables capture common failure modes for community notes.
Finally, $r_a$ is an auxiliary diagnostic signal measuring whether the note changes the reader's understanding of the post.

These dimensions take inspirations from community notes' rating process. In Community Note program, raters explain why a note is helpful or not helpful through predefined reason categories, including whether the note is clear, well-sourced, incorrect, unnecessary, etc.
MultiCom adopts this structure by eliciting diverse reason-level signals from simulated raters and using these features to augment aggregation. 
This design preserves preference nuances, where two notes may receive similar helpfulness votes while the other features such as source quality or claim coverage are different. 
In our representation, $y_a$ and $\mathbf{s}_a$ capture the agent's rating stance, $c_a$ captures confidence, $\mathbf{q}_a$ captures positive quality evidence, and $\mathbf{f}_a$ captures diagnostic failure modes. 


\subsection{Calibrated Multi-View Aggregation}

Obtaining structured outputs in JSON format from different persona agents, MultiCom clusters them into several note-level features, including raw vote distributions, confidence statistics, consistency signals, features regarding the reasons of helpfulness or not-helpfulness, cluster-level disagreement patterns, and features derived from the metadata~\cite{liu2023g,hashemi2024llm,ye2023flask}. A complete list of feature views and out-of-fold predictors is provided in Appendix~\ref{sec:appendix_aggregation_details}. These features enable the aggregator to model how agents vote and why they vote that way. 

We then use an out-of-fold method to process all learned aggregation components~\cite{wolpert1992stacked,kaufman2012leakage}. Specifically, for each individual note, the intermediate predictions used by the final aggregator are generated by models that were not trained using that particular note. This avoids over-fitting.

MultiCom finally integrates multiple complementary out-of-fold predictors, including weighted ensemble predictions~\cite{dietterich2000ensemble,caruana2004ensemble}, gated ensemble predictions, rescue-gate predictions~\cite{jacobs1991adaptive}, rationale blend predictions, and metadata predictions. For a note \(n\), each predictor \(m\) produces a label \(\hat{y}_{m,n} \in \{\texttt{NH}, \texttt{NMR}, \texttt{H}\}\). The final class score is calculated as

\begin{equation}
    S_c(n)=\sum_m w_m \mathbb{I}(\hat{y}_{m,n}=c),
\end{equation}

where \(w_m\) is the weight assigned to predictor \(m\). The final prediction is $\hat{y}_n = \arg\max_c S_c(n).$

Furthermore, in instances where the initial prediction from the ensemble model is \texttt{NMR}, we employ a conservative upgrading rule.
Specifically, if two auxiliary out-of-fold predictors consistently predict the same resolved label (i.e., \texttt{H} or \texttt{NH}), and the diagnostic statistics at the voting level satisfy a preset threshold, we upgrade the prediction result from \texttt{NMR} to that resolved label.
Detailed information regarding the auxiliary predictors, their input, and the upgrading thresholds are in Appendix~\ref{sec:appendix_aggregation_details}.

\section{Experiments}

\subsection{Methods}

We compare methods representing direct helpfulness prediction. We exclude other multi-agent systems~\cite{wu2024autogen,park2023generative} due to their similarity to MultiCom's ablation settings, and omit fact-checking systems~\cite{wang2024factcheck} as their tasks diverge from helpfulness prediction:

$\bullet$ Single Agent: 
This setting explores whether one agent alone is sufficient to substitute the multi-agent simulation process.
It receives the same post and note as MultiCom, and is tasked with generating a set of ratings similar to the agent in MultiCom, including overall helpfulness status, confidence, agreement signals, helpfulness reasons, not-helpfulness reasons, and diagnostic signals. 
Unlike MultiCom, this baseline employs no persona prompts derived from evaluator. 
We fed the outputs generated by this single agent into an out-of-fold calibration model, identical to those of MultiCom, ensuring fair comparisons.

$\bullet$ Fine-tuned Model:
Following prior work~\cite{xing2026communitynotes}, we use Mistral-7B-Instruct-v0.3 as the backbone model and fine-tune it with LoRA~\cite{hu2022lora,nguyen2026parameter,xing2026communitynotes} on the ComRate training set. 
Following Xing et al.~\cite{xing2026communitynotes}, each input instance consists of the post text, the community note text, and a classification instruction (predict \texttt{H}/\texttt{NH}/\texttt{NMR}). 
Additional implementation details are provided in Appendix~\ref{sec:appendix_finetune_details}.


$\bullet$ MultiCom: this setting is detailed in Sec~\ref{sec:multicom}.

\begin{table*}[!htbp]
    \centering
    \resizebox{\textwidth}{!}{%
    \begin{tabular}{ll cccc cccc}
        \toprule
        \textbf{Method} & \textbf{Metric}
        & \begin{tabular}[c]{@{}c@{}}Manipulated\\ media\end{tabular}
        & \begin{tabular}[c]{@{}c@{}}Factual\\ error\end{tabular}
        & \begin{tabular}[c]{@{}c@{}}Outdated\\ information\end{tabular}
        & \begin{tabular}[c]{@{}c@{}}Missing\\ important context\end{tabular}
        & \begin{tabular}[c]{@{}c@{}}Unverified\\ claim as fact\end{tabular}
        & Satire
        & \begin{tabular}[c]{@{}c@{}}Not\\ misleading\end{tabular}
        & Avg. \\
        \midrule

        \multirow{3}{*}{Single Agent}
        & Acc. & 69.7\%$_{3.8}$ & 79.8\%$_{1.4}$ & 83.4\%$_{2.0}$ & 81.4\%$_{1.3}$ & 81.4\%$_{1.7}$ & 74.7\%$_{4.4}$ & 85.7\%$_{1.6}$ & 80.9\%$_{0.9}$ \\
        & Bal. Acc. & 33.2\% & 38.0\% & 38.6\% & 37.4\% & 35.7\% & 35.3\% & 50.5\% & 38.6\% \\
        & Mac. F1 & 32.8\% & 38.1\% & 39.1\% & 37.7\% & 35.5\% & 35.8\% & 33.5\% & 38.6\% \\

        \cmidrule(lr){2-10} 
        \multirow{3}{*}{Fine-tuned Model}
& Acc. & 57.9\%$_{4.1}$ & 62.4\%$_{1.7}$ & 52.1\%$_{5.1}$ & 66.2\%$_{2.6}$ & 75.9\%$_{5.8}$ & 43.8\%$_{12.4}$ & 77.3\%$_{1.9}$ & 65.3\%$_{1.1}$ \\
& Bal. Acc. & 44.4\% & 35.4\% & 51.2\% & 32.0\% & 49.6\% & 17.9\% & 44.4\% & 35.3\% \\
& Mac. F1 & 34.7\% & 32.2\% & 33.9\% & 31.4\% & 43.5\% & 21.2\% & 47.6\% & 32.8\% \\

        \midrule

        \multirow{3}{*}{MultiCom w/o Cluster}
        & Acc. & 67.6\%$_{3.9}$ & 79.2\%$_{1.4}$ & 79.2\%$_{2.2}$ &  80.0\%$_{1.3}$ & 78.2\%$_{1.8}$ & 76.8\%$_{4.2}$ & 83.6\%$_{1.7}$ & 80.5\%$_{0.9}$ \\
        & Bal. Acc. & 38.2\% & 34.3\% & 38.6\% & 35.6\% & 34.8\% & 38.7\% & 56.1\% & 38.8\% \\
        & Mac. F1 & 39.1\% & 34.3\% & 38.2\% & 35.6\% & 34.7\% & 40.0\% & 35.4\% & 38.3\% \\
        \cmidrule(lr){2-10} 
        \multirow{3}{*}{MultiCom w/o MultiDim}
        & Acc. & 62.8\%$_{4.0}$ & 67.5\%$_{1.6}$ & 69.0\%$_{2.5}$ & 68.4\%$_{1.5}$ & 63.3\%$_{2.1}$ & 59.6\%$_{4.9}$ & 37.4\%$_{2.2}$ & 60.4\%$_{1.1}$ \\
        & Bal. Acc. & 43.1\% & 45.4\% & 51.7\% & 38.7\% & 39.9\% & 54.9\% & 58.4\% & 46.1\% \\
        & Mac. F1 & 32.6\% & 37.3\% & 38.5\% & 34.0\% & 32.4\% & 35.0\% & 21.5\% & 34.6\% \\

        \cmidrule(lr){2-10} 
        \multirow{3}{*}{\textbf{MultiCom}}
        & Acc. & 76.6\%$_{3.5}$ & 84.1\%$_{1.2}$ & 83.1\%$_{2.1}$ & 84.1\%$_{1.2}$ & 83.2\%$_{1.6}$ & 76.8\%$_{4.2}$ & 90.9\%$_{1.3}$ & 84.7\%$_{0.8}$ \\
        & Bal. Acc. & 57.3\% & 66.5\% & 65.6\% & 66.4\% & 64.6\% & 69.9\% & 78.4\% & 68.3\% \\
        & Mac. F1 & 52.9\% & 59.9\% & 58.2\% & 59.6\% & 58.1\% & 60.4\% & 46.7\% & 60.1\% \\

        \bottomrule
    \end{tabular}
    }%
    \caption{Performance comparison (accuracy, balanced accuracy, and Macro-F1) across different methods on the ComRate dataset. Subscripts denote binomial standard errors~\cite{basharat2025variantbench}.}
    \label{tab:main_experiment}
\end{table*}

\subsection{Evaluation Process}

We evaluated all methods in the ComRate evaluation set, containing 2,000 samples, where we applied stratified sampling based on the notes' creation year, topic, and final status. Each instance includes a post, a community note, and a ground-truth label corresponding to its official status: \texttt{H}, \texttt{NH}, or \texttt{NMR}. The \texttt{NMR} category is the majority class, representing 88.75\% of the evaluation dataset.

All trainable components, including fine-tuned models and calibrated aggregation models, were evaluated using a 5-fold stratified out-of-fold methodology. In each fold, 80\% of data was used for training, and the remaining 20\% served as the held-out test set. Final metrics were calculated by aggregated predictions across all five held-out sets, ensuring no model was tested on its training data.

Given the class imbalance, we evaluate performance using accuracy, balanced accuracy ($\frac{1}{3}\sum_{i=1}^{3}\text{Recall}_{i}$) and macro-F1 ($\frac{1}{3}\sum_{i=1}^{3}\text{F1}_{i}$).

\subsection{Parameters}

In MultiCom, the number of agents determines the volume of simulated judgments. Our primary setup uses 16 persona agents, each mapping to a distinct rater cluster derived from matrix factorization. When scaling to 32 or 48 agents, we retain the original 16 cluster profiles and assign multiple independent replicas to each. The replicas mimics randomness of each persona agent, which improves rating diversity.
For calibrated aggregation, we use nested cross-validation. 
The outer loop uses 5-fold stratified cross-validation for final evaluation. 
Within each outer training split, we used inner 5-fold cross-validation to set logistic regularization strength, class weights, and decision thresholds~\cite{adekoya2025ensemble,stenhouse2021development}. 


\section{Results}

\subsection{Main Study}

Table~\ref{tab:main_experiment} summarizes the performance. MultiCom achieves the best balanced accuracy of 68.3\% and Macro-F1 of 60.1\%, while maintaining high overall accuracy of 84.7\%. 
The fine-tuned LoRA baseline performs substantially worse than MultiCom, especially in balanced accuracy and Macro-F1, indicating that direct text classification struggles with the three-way helpfulness task under severe class imbalance.

A category-level analysis reveals that the fine-tuned model has lower accuracy on specific categories, such as ``Satire'' (43.8\%) and ``Outdated information'' (52.1\%).
This suggests the limitations of relying solely on text classification for community note evaluation without crowdsourced reasoning.
Conversely, MultiCom maintains stronger and more balanced performance across diverse factual contexts, achieving high accuracy on ``Not misleading'' (90.9\%), ``Factual error'' (84.1\%), and ``Missing important context'' notes (84.1\%).
To address class imbalance, we further constructed a balanced evaluation set of 1,998 notes (666 per class) while preserving the natural distribution of other features. On this balanced set, MultiCom consistently outperforms alternative methods. For instance, MultiCom achieves an accuracy, balanced accuracy, and macro-F1 of 68.0\%, 68.0\%, and 67.3\%, whereas the single-agent baseline only yields 49.3\%, 49.2\%, and 46.2\%, respectively (see Appendix~\ref{app:balanced_set} for details).

\subsection{Ablation Study}

We conduct ablation study to compare MultiCom with two ablated variants: \textit{MultiCom w/o Cluster}, removing cluster-grounded persona profiles while preserving the structured output schema, and \textit{MultiCom w/o MultiDim}, removing multi-dimensional diagnostic signals and relying only on agents' helpfulness votes.

As in Table~\ref{tab:main_experiment}, MultiCom achieves the best overall performance, with an average accuracy of 84.7\%, balanced accuracy of 68.3\%, and Macro-F1 of 60.1\%. 
Removing rater-grounded clustering reduces average accuracy from 84.7\% to 80.5\%, while balanced accuracy and Macro-F1 drop sharply from 68.3\% to 38.8\% and from 60.1\% to 38.3\%, respectively.

Conversely, removing multi-dimensional diagnostic mechanism leads to more severe perfomance degradation. \textit{MultiCom w/o MultiDim} achieves only 60.4\% average accuracy, 46.1\% balanced accuracy, and 34.6\% Macro-F1. 
Compared to MultiCom, these metrics dropped by 24.3, 22.2, and 25.5 percentage respectively. 
The performance decline is particularly acute for categories such as \textit{Not misleading}, where accuracy decreases from 90.9\% to 37.4\%. 
These results show that diagnostic dimensions, such as evidence quality and claim coverage, provide important signals for classification.

\subsection{Generalizability Study}

As detailed in Table~\ref{tab:generalizability}, MultiCom showed generalizability across models and temporal aspects. For models, we use the same evaluation set comprising 2000 notes and varied backbone models (e.g., qwen, claude). Regarding temporal aspects, we employ a rolling future-prediction method, predicting the future 1-3 years' note helpfulness based on prior 1-3 years' windows (see Appendix~\ref{app:generalizability} for details and justification).
All feasible temporal windows are averaged when reporting accuracies.
MultiCom achieved peak average accuracies of 73.1\% and 71.5\% using DeepSeek-v3.2 and GPT-5.1. In multi-year settings, it has accuracies between 84.9\% and 87.1\% across one- to three-year forecasting windows, indicating minimal performance degradation over time. 

\begin{table*}[!htbp]
    \centering
    \resizebox{\textwidth}{!}{%
    \begin{tabular}{l cccc cccc} 
        \toprule
         & \begin{tabular}[c]{@{}c@{}}Manipulated\\ media\end{tabular} & \begin{tabular}[c]{@{}c@{}}Factual\\ error\end{tabular} & \begin{tabular}[c]{@{}c@{}}Outdated\\ information\end{tabular} & \begin{tabular}[c]{@{}c@{}}Missing\\ important context\end{tabular} & \begin{tabular}[c]{@{}c@{}}Unverified\\ claim as fact\end{tabular} & Satire & \begin{tabular}[c]{@{}c@{}}Not\\ misleading\end{tabular} & Avg. \\\midrule
        GPT-4o & 49.0\%$_{4.2}$ & 64.0\%$_{1.7}$ & 71.9\%$_{4.6}$ & 61.1\%$_{2.7}$ & 64.8\%$_{6.5}$ & 43.8\%$_{12.4}$ & 83.3\%$_{1.7}$ & 67.4\%$_{1.0}$ \\
        GPT-5.1 & 55.9\%$_{4.1}$ & 68.7\%$_{1.6}$ & 71.9\%$_{4.6}$ & 69.7\%$_{2.5}$ & 70.4\%$_{6.2}$ & 75.0\%$_{10.8}$ & 80.4\%$_{1.8}$ & 71.5\%$_{1.0}$ \\
        Claude Opus 4.6 & 49.7\%$_{4.2}$ & 60.8\%$_{1.7}$ & 66.3\%$_{4.8}$ & 60.5\%$_{2.7}$ & 67.9\%$_{6.4}$ & 50.0\%$_{12.5}$ & 71.7\%$_{2.1}$ & 63.4\%$_{1.1}$ \\
        Qwen3-Max & 39.3\%$_{4.1}$ & 54.9\%$_{1.8}$ & 59.4\%$_{5.0}$ & 57.6\%$_{2.7}$ & 51.9\%$_{6.8}$ & 56.2\%$_{12.4}$ & 77.5\%$_{1.9}$ & 59.9\%$_{1.1}$ \\
        DeepSeek-v3.2 & 52.1\%$_{4.2}$ & 72.8\%$_{1.6}$ & 66.7\%$_{4.8}$ & 69.4\%$_{2.5}$ & 74.1\%$_{6.0}$ & 56.2\%$_{12.4}$ & 84.0\%$_{1.7}$ & 73.1\%$_{1.0}$ \\
        \midrule
        Pred Year & 80.5\%$_{3.5}$ & 86.3\%$_{1.3}$ & 89.0\%$_{3.5}$ & 87.2\%$_{2.0}$ & 83.3\%$_{5.8}$ & 76.9\%$_{11.7}$ & 90.8\%$_{1.4}$ & 87.1\%$_{0.8}$ \\
        Pred 2Years & 72.9\%$_{3.3}$ & 83.5\%$_{1.2}$ & 86.1\%$_{3.2}$ & 86.2\%$_{1.8}$ & 83.1\%$_{4.9}$ & 66.7\%$_{11.1}$ & 90.5\%$_{1.3}$ & 84.9\%$_{0.7}$ \\
        Pred 3Years & 73.6\%$_{3.1}$ & 83.8\%$_{1.1}$ & 87.3\%$_{3.0}$ & 86.6\%$_{1.7}$ & 82.8\%$_{4.7}$ & 78.9\%$_{9.4}$ & 90.3\%$_{1.2}$ & 85.0\%$_{0.7}$ \\

        \bottomrule
    \end{tabular}
    }%
    \caption{Accuracy across different generalizability settings. Underscripts denoted binomial standard errors.}
    \label{tab:generalizability}
\end{table*}

\begin{figure}[!htbp]
    \centering
    \includegraphics[width=0.96\linewidth]{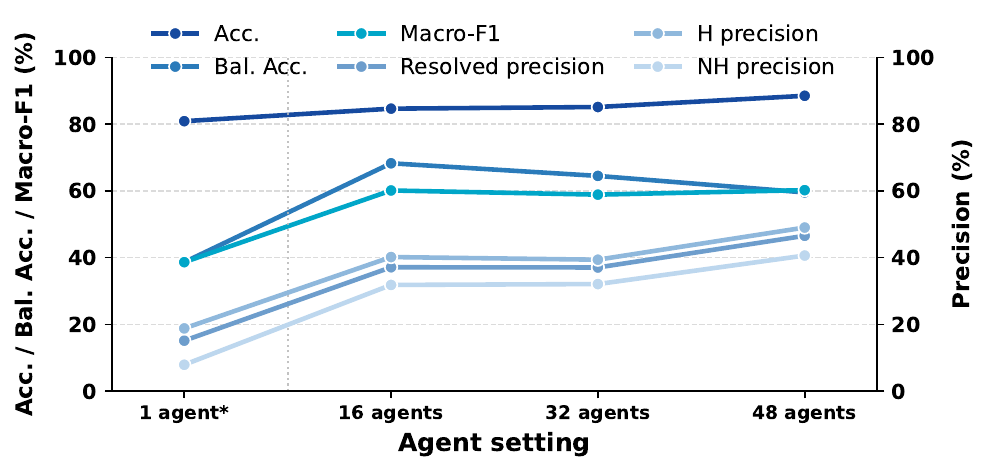}
    \caption{Accuracy and precision across different agent numbers, where ``1 agent*'' is the baseline without persona simulation. 
}
    \label{fig:agent_panel_generalization}
\end{figure}

Figure~\ref{fig:agent_panel_generalization} evaluates whether MultiCom remains effective with different numbers of persona agents.
Here, we retain the same set of 16 cluster personality profiles, assigning multiple independent agents to each cluster, 2 for 32-agent configuration and 3 for 48-agent configuration. Agents belonging to the same cluster share an identical cluster-specific prompt, yet processing queries independently. This setting aims to assess the impact of conducting repeated, independent simulations within each evaluation group.
Results show that the 32-agent and 48-agent settings achieve higher overall accuracy than the 16-agent setting. 
However, balanced accuracy degraded, where more agents cause MultiCom to behave conservatively when handling the \texttt{NMR} category, reducing recall for \texttt{H} or \texttt{NH}. 

\subsection{Agent-Human Alignment}

\begin{figure}[!htbp]
    \centering 
    \includegraphics[width=0.50\textwidth]{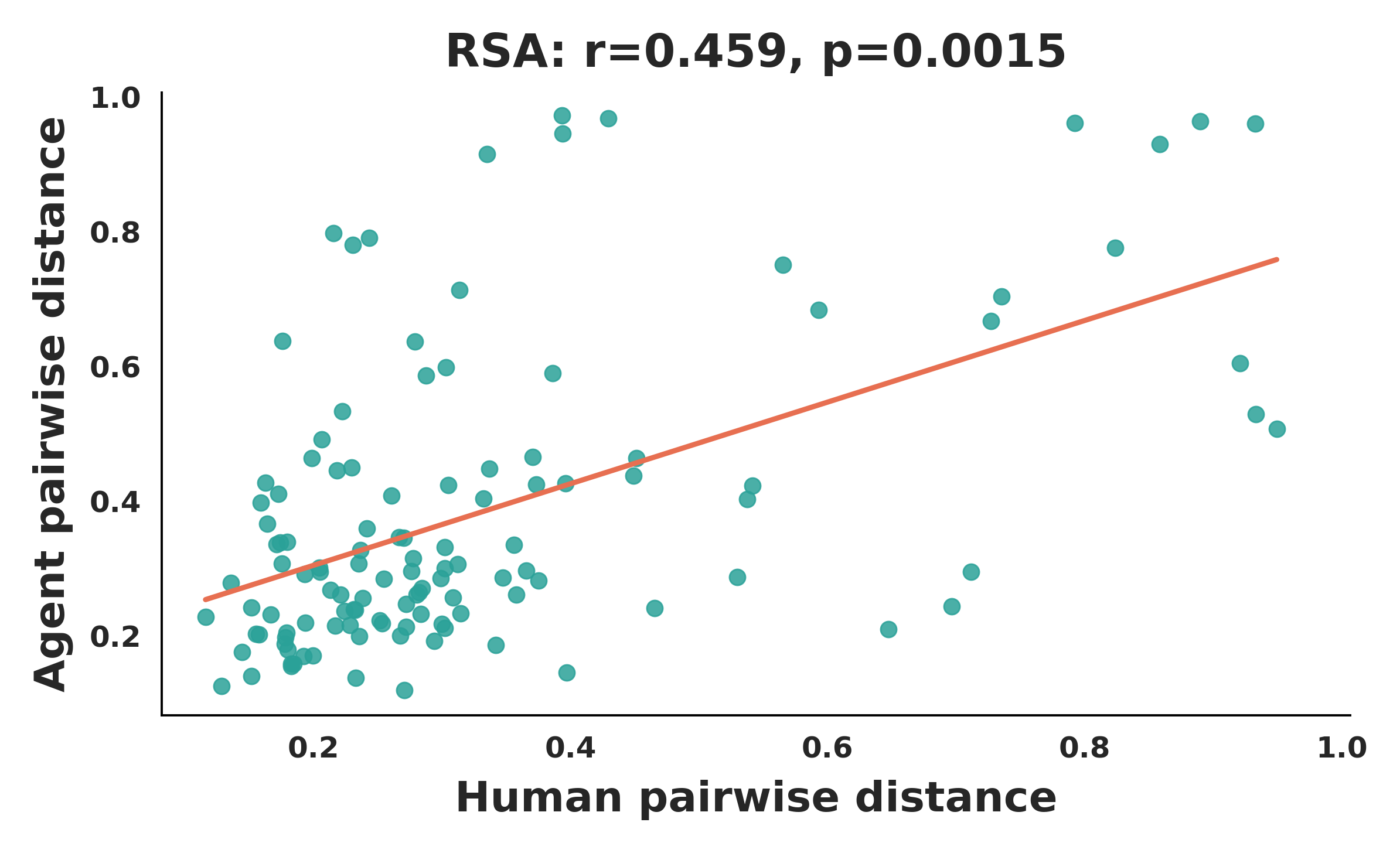}
    \caption{Representational alignment between humans' ratings and agents' ratings.}
    \label{fig:alignment}
\end{figure}

To evaluate structural alignment between agent and human rating patterns, we conducted a Representational Similarity Analysis (RSA) across the 16 clusters, a standard paradigm for human-AI alignment evaluation~\cite{sucholutsky2025getting,wynn2024learning}. Representational Dissimilarity Matrices (RDMs) for both humans and agents were constructed by computing the absolute difference between mean note-level scores for each cluster pair, restricted to mutually evaluated notes. We then calculated RSA score as Spearman correlation between the upper-triangular entries of the aligned RDMs. A label-permutation test (2,000 iterations) confirmed significant structural alignment ($r = 0.459$, $p = .0015$, see Figure~\ref{fig:alignment}). Additionally, we  tested its generalizability across all models, finding consistently high RSA across all models (e.g., claude: $r = 0.714$, $p = .0005$; qwen: $r = 0.329$, $p = .0135$). We further tested its generalizability across different temporal aspects, finding that agent simulations exhibit consistent structures with human clusters even for predicting future data (Pred year: $r = 0.376$, $p = .0015$; Pred 2Years: $r = 0.376$, $p = .002$; Pred 3Years: $r = 0.376$, $p = .0015$; see Appendix~\ref{app:representational} for details).

\subsection{Computational Cost and Latency}

MultiCom's computational cost comes from LLM-based agent simulation. 
The aggregation stage requires no additional LLM usage, and only involves feature construction and logistic-regression inference.
The price of GPT-5.4-nano is \$0.20/\$1.25 per million input/output tokens. Therefore, on average MultiCom costs \$0.0076 for each note. 
Across all models we benchmarked, MultiCom costs between \$0.0075 and \$0.170 per note, suggesting its economical viability.

\subsection{Robustness on Instable Notes} 
We further demonstrated MultiCom's robustness through examining notes with status-transition signals and notes with or without author-indicated trustworthy sources. 
Status-transition notes obtain lower accuracy than the full evaluation set (75.7\% vs. 84.7\%), suggesting that volatile notes are more ambiguous. 
The trustworthy-source split shows stability of MultiCom even for notes without trustworthy sources, which shows even higher accuracy (87.8\%). 
Full results are provided in Appendix~\ref{sec:appendix_additional_robustness}.

\subsection{Effects of NMR Ratio}

\begin{figure}[!htbp]
    \centering 
    \includegraphics[width=0.50\textwidth]{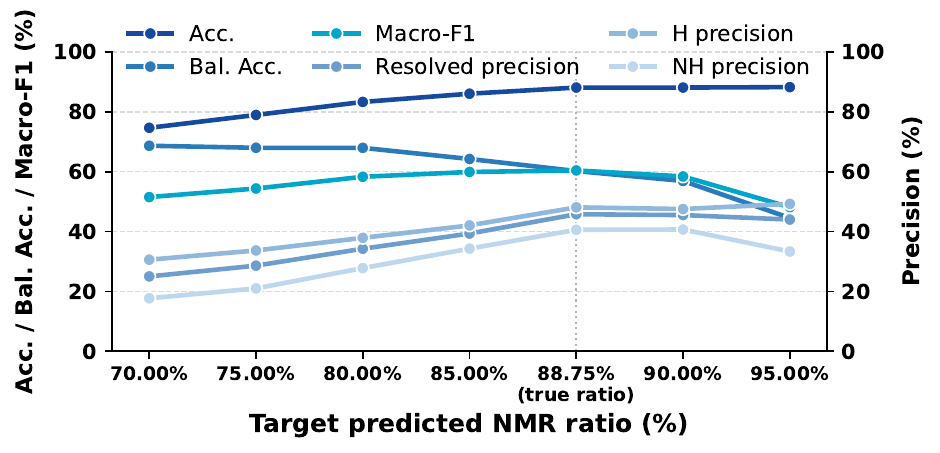}
    \caption{Accuracy and resolved-class precision across varied predicted NMR ratios.}
    \label{fig:need_more_rating}
\end{figure}

Because \texttt{NMR} is the majority class in ComRate, we explore how the predicted \texttt{NMR} ratio affects MultiCom's performance by varying the decision threshold. The ground-truth ratio in the dataset is 88.75\%. 
As depicted in Figure~\ref{fig:need_more_rating}, we found a trade-off between balanced accuracy, and precision for \texttt{H} and \texttt{NH} classes. Increasing the predicted \texttt{NMR} ratio improves overall accuracy due to class imbalance, and enhances the precision of \texttt{H} and \texttt{NH} but at the cost of balanced accuracy. Conversely, lowering the ratio toward 70.00\% improves balanced accuracy but diminishes precision of \texttt{H} and \texttt{NH}. This suggests that \texttt{NMR} threshold could function as an adjustment parameter, where conservative abstention lead to more reliable predictions.

\begin{table*}[htbp]
    \centering
    \resizebox{\textwidth}{!}{%
    \begin{tabular}{l cccc cccc}
        \toprule
        \multirow{2}{*}{Reason label / Metric} 
        & \multicolumn{4}{c}{All notes}
        & \multicolumn{4}{c}{Correct H/NH notes} \\
        \cmidrule(lr){2-5} \cmidrule(lr){6-9}
        & Prevalence & Precision & Recall & F1
        & Prevalence & Precision & Recall & F1 \\
        \midrule
        \texttt{helpfulClear} 
        & 64.2\% & 64.4\% & 99.7\% & 78.2\%
        & 61.9\% & 62.9\% & 100.0\% & 77.2\% \\
        
        \texttt{helpfulAddressesClaim}
        & 59.4\% & 59.4\% & 100.0\% & 74.5\%
        & 53.7\% & 53.7\% & 100.0\% & 69.9\% \\
        
        \texttt{helpfulImportantContext}
        & 54.3\% & 54.4\% & 99.9\% & 70.4\%
        & 56.7\% & 56.7\% & 100.0\% & 72.4\% \\
        
        \texttt{helpfulGoodSources}
        & 24.1\% & 30.8\% & 75.5\% & 43.7\%
        & 33.6\% & 40.2\% & 77.8\% & 53.0\% \\
        
        \texttt{notHelpfulMissingKeyPoints}
        & 17.3\% & 18.4\% & 84.1\% & 30.2\%
        & 24.6\% & 25.0\% & 78.8\% & 38.0\% \\
        
        \texttt{notHelpfulSourcesMissingOrUnreliable}
        & 7.0\% & 12.1\% & 37.1\% & 18.2\%
        & 14.2\% & 25.8\% & 42.1\% & 32.0\% \\
        
        \texttt{notHelpfulArgumentativeOrBiased}
        & 10.5\% & 14.6\% & 58.8\% & 23.4\%
        & 17.2\% & 35.6\% & 69.6\% & 47.1\% \\
        
        \texttt{notHelpfulNoteNotNeeded}
        & 14.5\% & 16.3\% & 85.2\% & 27.4\%
        & 14.9\% & 14.2\% & 75.0\% & 23.8\% \\
        
        \texttt{notHelpfulIncorrect}
        & 8.7\% & 11.7\% & 36.2\% & 17.7\%
        & 10.4\% & 10.0\% & 21.4\% & 13.6\% \\
        
        \midrule
        \textbf{Overall F1} 
        & \multicolumn{4}{c}{\textit{Micro: 54.3\%} \quad \textit{Macro: 42.6\%} \quad \textit{Sample: 53.3\%}} 
        & \multicolumn{4}{c}{\textit{Micro: 56.2\%} \quad \textit{Macro: 47.4\%} \quad \textit{Sample: 55.9\%}} \\
        \bottomrule
    \end{tabular}
    }
    \caption{Per-label and overall multi-label prediction performance for community notes reason labels. The H/NH subset includes only notes whose Helpful/Not Helpful status is correctly predicted by MultiCom.}
    \label{tab:reason_prediction}
\end{table*}

\subsection{Fine-grained Reasons Prediction}

To provide fine-grained prediction rationales, similar to Xing et al.~\cite{xing2026communitynotes}, we examined accuracy of predicting fine-grained reason labels, including four \texttt{H} labels and five \texttt{NH} labels adopted from the Community Notes program (details in Appendix~\ref{app:label}). 
Table~\ref{tab:reason_prediction} reports the reason prediction's results. Features extracted by MultiCom achieve a Micro-F1, Macro-F1, and Sample-F1 of 54.3\%, 42.6\%, and 53.3\% respectively. We further report results where MultiCom correctly predicts the \texttt{H}/\texttt{NH} statuses, thereby isolating label prediction errors. In this subset, Micro-F1 increases to 56.2\% while Sample-F1 reaches 55.9\%. Among various reason-level metrics, those high-frequency dimensions are most accurate, such as \texttt{Clear}, \texttt{AddressesClaim}, and \texttt{ImportantContext}, indicating MultiCom's effectiveness for status prediction.

\section{Discussion}

\textbf{Grounded simulation.} Single agent evaluations often collapse into artificial consensus, especially for evaluation needing multiple perspectives. MultiCom anchors persona agents in a matrix-factorized space derived from real-world rating behaviors, preserving the genuine heterogeneity and ideological disagreements of human raters.


\textbf{Trade-offs.} As analyzed, the predicted NMR ratio could serve as a lever, where platforms could dynamically balance the trade-off between abstention and H/NH recall. For example, platforms can enforce stricter thresholds during breaking news events to avoid premature judgments.

\textbf{Explainability.} Beyond binary classification, MultiCom yields actionable diagnostic feedback detailing specific failure modes and evidence quality. This transparency supports human-in-the-loop systems, allowing users to verify the underlying reasoning prior to formalized consensus.

\textbf{Applications.} MultiCom could be used to evaluate note helpfulness for platform deployment, and guiding contributors in revisions via multi-dimensional feedback. These interpretable scores can assist novice fact-checkers and function as reward signals for AI-generated notes, such as for the $\mathbb{X}$'s ``AI Note Writer'' or ``Collaborative Note''.

\section{Related Work}

\textbf{Community-based fact-checking.} The rapid online spread of misinformation has become a critical societal concern~\cite{scheufele2019science}. In response, community-based fact-checking emerged, leveraging the crowd's intelligence for detecting misinformation~\cite{prollochs2022community,borenstein2025can}, and writing structured notes to deter the spreading of misinformation~\cite{chuai2024did}. They are shown to be highly effective in countering misinformation~\cite{chuai2026community}, while also be known to have resilience issues~\cite{chuai2026consensus}. Many recent initiatives began to use AI to automate the community-based fact-checking processes~\cite{de2025supernotes,zhang2025commenotes,zhang2026collab}, especially in specific domains~\cite{wu2025beyond}. 

\textbf{Helpfulness evaluation.} Verifying the accuracy of fact-checks is essential~\cite{wang2024factcheck}. This could be achieved through cross-checking references~\cite{smeros2021sciclops}. For fact-checkers, this also involve scrutinization among organizations~\cite{warren2025show,juneja2022human}. More importantly, because debunking texts authored by crowds or AI can be inconsistent, evaluating their helpfulness remains more crucial for improving fact-checking capabilities~\cite{nakov2021automated}. While early work proposed fine-tuning models to assess community notes' helpfulness, they overlook the \texttt{NMR} status, and ignore notes' temporal dynamics, where newer notes contain updated context unavailable to older ones.

\section{Conclusion}

This paper introduces MultiCom, a persona-guided multi-agent framework designed to evaluate crowdsourced debunking notes. We construct ComRate, a large-scale, real-world rating dataset based on $\mathbb{X}$'s Community Notes. By modeling empirical behavioral heterogeneity and eliciting multi-dimensional diagnostic judgments, MultiCom accurately simulates human rater consensus. Extensive evaluations show that our framework significantly outperforms alternative baselines, providing a scalable, accurate, and explainable solution for automated content governance.

\section{Limitations}

This paper conducted experiments primarily around the Community Note dataset. The effectiveness of multi-agent simulations has not been validated across other content governance platforms (e.g., Meta, YouTube), which may possess different user demographics, moderation architectures, and interface constraints. Besides, the current dataset and evaluation pipeline are primarily centered on English-language interactions and Western-centric misinformation contexts. Future work could expand the scope to general fact-checking datasets and contexts. 

\section{Ethical considerations}

This paper uses the ComRate dataset, which was constructed using publicly available data, or data from the $\mathbb{X}$ API. The data collection process strictly adheres to the platform's terms of service and data use guidelines. Because the official Community Notes ecosystem anonymizes contributor identities by design, our dataset and subsequent multi-agent simulations do not compromise personally identifiable information or individual user privacy. 

A potential ethical consideration in automated fact-checking is the potential for algorithmic bias. To mitigate the risk of artificial consensus often found in single-agent evaluations, MultiCom anchors its persona agents in a matrix-factorized space derived from real-world rating behaviors. While this design intentionally preserve the heterogeneity and nuanced ideological disagreements of human raters, we acknowledge that the underlying LLMs driving the agents may still have biases from their pre-training corpora. To counterbalance this, MultiCom relies on multi-dimensional diagnostics judgments instead of binary classifications. This potentially improves evidence quality, and enabling investigation of potential failures modes.

Finally, we emphasized the potential dual-use nature of this framework. MultiCom is designed to support, instead of replacing the human-driven content moderation. It could empower human users to verify the simulated reasoning prior to reaching a formalized consensus, instead of discrediting specific notes, or fact-checkers. By providing these feedback such as helpfulness ratings and reasons, MultiCom is developed to foster a healthier information ecosystem.



\bibliography{main}

\appendix

\clearpage

\section{Generative AI Usage}

In accordance with generative AI usage policies, we disclose the use of Generative AI tools. We used generative AI as the base model for the experiment. Besides, we utilized Google's Gemini 3 Pro and ChatGPT (i.e., GPT-5.2) as writing assistants. Its functions were limited to proofreading, language and clarity enhancement, conciseness, and word choice, and was not used to generate any core scientific content. Authors hold full responsibility to the paper's content.

\section{Dataset Details}

\subsection{Evaluation-Set Sampling Details}
\label{sec:appendix_sampling}

We sampled evaluation dataset from the ComRate dataset, applying the following inclusion criteria: (1) having a valid note status, (2) non-empty note text, (3) a valid corresponding post ID, and (4) available associated post text. The primary evaluation set consists of 2,000 notes, sampled proportionally based on creation year, final status, and primary note category. This approach aims to preserve the real-world class distribution characteristics of the ComRate dataset. Specifically, this set includes 1,775 \texttt{NMR}, 149 \texttt{H}, and 76 \texttt{NH}. 

To enhance robustness, we additionally constructed a balanced dataset consisting of 1,998 annotations, wherein each class (e.g., \texttt{NMR}, \texttt{H}, \texttt{NH}) contains 666 annotations. Within each category of this balanced dataset, annotation samples were drawn proportionally based on creation year and primary note category to reflect diversity in both temporal and thematic distribution.

In our evaluation process, we employed three non-overlapping data splits: a training set used to construct personality profiles, a primary evaluation set comprising 2,000 notes, and a balanced robustness evaluation set comprising 1,998 notes.
The training set used for constructing personality profiles consists of historical Community Notes rating data. Its purpose is to estimate the matrix-factorized rater space, and to derive generalized personality profiles for each clustering level.
Both of the aforementioned evaluation sets are entirely independent of the training set used for personality profile construction, and are used only for out-of-fold aggregation, model comparison, and robustness evaluation.

\subsection{Details for the Dataset Analysis}\label{sec:details_dataset}

Additional details are provided for the descriptive analyses reported in Figure~\ref{fig:analysis}. These analyses characterize ComRate from three aspects: temporal growth, rater heterogeneity, and note-level distributional patterns.

For the temporal analysis in Figure~\ref{fig:temporal_trend}, we aggregate the dataset by year. 
Each year, we compile statistics on the unique number of community notes, the unique number of posts associated with these notes, and the total count of rating records.
The statistics for 2026 are calculated using data available up to April 5, 2026, and therefore represent only a partial year's data.
Analysis show that the Community Notes program has expanded quickly, particularly since 2023 when both the volume of notes and ratings grew quickly.

For the rater-cluster analysis in Figure~\ref{fig:cluster_insight}, we used contributor representations learned from the biased rank-one matrix factorization model described in Section~\ref{sec:multicom}. 
Each observed rating is mapped to a numerical helpfulness value, from which the model estimates contributor-specific intercepts and latent factors within the contributor-note rating matrix. 
The contributor intercepts reflect the rater's overall leniency or strictness, while the latent factors capture residual variation in rating behavior after accounting for global and note-level effects.
We clustered the contributors into 16 rater groups, corresponding to the 16 role agents employed in the MultiCom setting.
For each group, we summarized both matrix-factorization features and interpretable behavioral statistics.
Specifically, the heatmap includes seven cluster-level statistics: 

$\bullet$ \textit{Rater intercept}: the average contributor bias term from the biased matrix-factorization model.

$\bullet$ \textit{Latent factor}: the average contributor latent coordinate.

$\bullet$ \textit{Agreement}: the average agreement tendency of contributors in the cluster.

$\bullet$ \textit{Mean note score}: the average score of notes rated by contributors in the cluster.

$\bullet$ \textit{Helpful share} and \textit{not-helpful share}: the average proportions of ratings marked helpful and not helpful.

$\bullet$ \textit{Notes authored}: the average note-authoring activity of contributors in the cluster.

Figure~\ref{fig:cluster_insight} visualizes these group-level statistics after feature-wise standardization. 
For each feature \(k\), let \(x_{c,k}\) denote the raw cluster-level value of feature \(k\) for cluster \(c\), and then standardize it across the 16 clusters:
\[
z_{c,k}=\frac{x_{c,k}-\bar{x}_k}{s_k},
\]
where \(\bar{x}_k\) and \(s_k\) are the mean and standard deviation of feature \(k\) across clusters. In the heatmap, each row corresponds to a rater cluster, each column corresponds to one behavioral statistic, and the color indicates the standardized value of the specific cluster's statistics.
Red cells indicate higher values, while blue cells indicate lower values.

The results indicate that the raters are heterogeneous.
Some groups are more inclined to rate helpful, while others are stricter. There are also differences in consistency, or rating numbers.
These observed differences provide an empirical basis for MultiCom's role-guided design: the simulated agents represent distinct rating behavior.

For Figure~\ref{fig:full_dataset_distributions}, we computed (1) distribution of note categories, and (2) notes per post. 
First, each note was assigned to primary misleading / not misleading categories based on its metadata fields. Note that one note could have multiple corresponding categories. For each category, we calculated the number of notes it contained as well as its corresponding proportion of the total.
Second, we calculated the ``note-to-post'' ratio by grouping notes according to their associated post IDs and counting the number of notes attached to each individual post.

\subsection{Additional Dataset Statistics}
\label{sec:appendix_dataset_statistics}

We provide additional descriptive statistics for the full ComRate note collection. 
These analyses are computed over all 2,566,644 notes in ComRate, rather than only the 2,000-note evaluation subset. 
Because the official Community Notes release does not provide complete post text for all notes, the full-dataset language/script and length statistics are computed from note text and official note metadata.

\begin{figure}[htbp]
    \centering
    \includegraphics[width=0.5\textwidth]{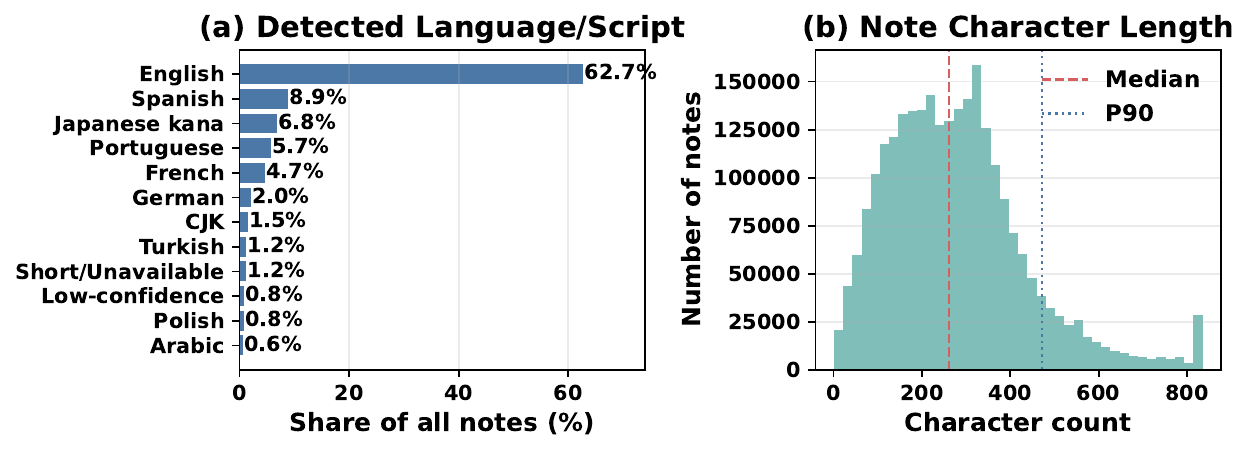}
    \caption{
    Distributions of note language/script and note characters length in ComRate.
    Panel (a) reports the dominant language or script category for each note.
    Panel (b) reports the characters length of note text, with dashed lines indicating the median and 90th percentile.
    }
    \label{fig:full_note_language_characters_distribution}
\end{figure}

Figure~\ref{fig:full_note_language_characters_distribution} summarizes the language/script and character-length distributions of the detected ComRate notes.
ComRate is predominantly composed of English notes, accounting for 62.75\% of all annotations. 
Other common Latin-script languages include Spanish (8.88\%), Portuguese (5.67\%), French (4.66\%), German (2.03\%), Turkish (1.24\%), and Polish (0.77\%).
Non-Latin scripts also appear in significant quantities, including Japanese (6.78\%), Chinese (1.47\%), and Arabic (0.63\%).

This distribution indicates that while ComRate is primarily English-centric, it also encompasses multilingual and cross-script examples drawn from the global Community Notes ecosystem, which increases the generalizability of our results.

The distribution of note lengths reveals that Community Notes are generally concise.
The average length of a note is 279.7 characters, with a median of 262 characters and a 90th percentile of 472 characters.
The 25th and 75th percentiles are 161 and 358 characters, suggesting that the majority of notes provide concise contextual explanations rather than lengthy fact-checking articles.


\section{Detailed Experiment Results}
\subsection{Robustness on Instable Notes}
\label{sec:appendix_additional_robustness}

We analyzed robustness on instable notes in evaluation sets~\cite{chuai2026consensus}, with results shown in Table~\ref{tab:appendix_status_source_robustness}. 
First, we identify notes with status-transition signals using the official status-history summary fields. 
A note is treated as a status-transition case if it has evidence of a previous resolved status that differs from the current status, or if the official status fields disagree across core, expansion, group, locked, current, first resolved, and most recent resolved statuses. 
This provides a conservative proxy for volatile notes whose public status may have changed over time. 
Second, we split notes by the author-provided \texttt{trustworthySources} field in the official note metadata.

\begin{table*}[htbp]
    \centering
    \resizebox{\textwidth}{!}{%
    \begin{tabular}{lrrrrrrrrrrr}
        \toprule
        Subset & $N$ & True NH & True NMR & True H & Acc. & Bal. Acc. & Macro-F1 & NH recall & NMR recall & H recall & Pred. NMR \\
        \midrule
        All notes & 2000 & 76 & 1775 & 149 & 84.7\% & 68.3\% & 60.1\% & 55.3\% & 87.8\% & 61.7\% & 1639 \\
        Status-transition notes & 169 & 20 & 116 & 33 & 75.7\% & 66.4\% & 66.4\% & 55.0\% & 83.6\% & 60.6\% & 115 \\
        Non-transition notes & 1831 & 56 & 1659 & 116 & 85.5\% & 68.5\% & 58.3\% & 55.4\% & 88.1\% & 62.1\% & 1524 \\
        Trustworthy source = yes & 1657 & 58 & 1454 & 145 & 84.0\% & 66.0\% & 59.6\% & 48.3\% & 87.6\% & 62.1\% & 1348 \\
        Trustworthy source = no & 343 & 18 & 321 & 4 & 87.8\% & 72.2\% & 59.0\% & 77.8\% & 88.8\% & 50.0\% & 291 \\
        \bottomrule
    \end{tabular}
    }
    \caption{MultiCom performance on status-transition and trustworthy-source subsets. ``Status-transition'' is derived from official status-history summary fields and should be interpreted as a proxy for volatile note status.}
    \label{tab:appendix_status_source_robustness}
\end{table*}

The status-transition subset has lower overall accuracy than the full evaluation set, suggesting that notes with unstable or changing official status are more difficult cases. 
However, its balanced accuracy remains close to the overall result because this subset contains a larger fraction of resolved \texttt{Helpful} and \texttt{Not Helpful} notes.

The trustworthy-source split shows that source metadata changes the difficulty profile rather than producing a simple monotonic effect. 
Notes with \texttt{trustworthySources}=1 form the majority of the evaluation set and include more resolved \texttt{Helpful} examples, while notes with \texttt{trustworthySources}=0 are more dominated by \texttt{Needs More Ratings}. 
MultiCom obtains higher overall accuracy on the no-trustworthy-source subset, largely because most of these examples are unresolved, but this subset contains very few \texttt{Helpful} notes, making resolved-class recall less stable.

\subsection{Representational Similarity Analysis}\label{app:representational}

We quantify structural alignment between agent and human rating behaviors at the cluster level using representational similarity analysis (RSA), a standard approach for comparing representational similarities across humans and AIs~\cite{sucholutsky2025getting,wynn2024learning}.  
Let $\mathcal{C}=\{0,\dots,15\}$ denote the 16 clusters. For each pair $(i,j)\in\mathcal{C}^2$, we define a shared note set
\[
S_{ij}=\mathcal{N}^{H}_i \cap \mathcal{N}^{H}_j \cap \mathcal{N}^{A}_i \cap \mathcal{N}^{A}_j,
\]
where $\mathcal{N}^{H}_k$ and $\mathcal{N}^{A}_k$ are notes rated by human cluster $k$ and agent cluster $k$, respectively. We retain a pair only if $|S_{ij}|\ge 30$ to ensure reliable distance estimates.

Human note-level ratings are encoded as $\{0,0.5,1\}$ for \texttt{not\_helpful}, \texttt{somewhat\_helpful}, and \texttt{helpful}. Agent note-level ratings use the corresponding scalar predicted helpfulness score. For each retained pair $(i,j)$, we compute dissimilarity as the absolute difference in cluster-wise mean score on the shared set:
\[
D^H_{ij}=\left|\mu^H_i(S_{ij})-\mu^H_j(S_{ij})\right|,\]

\[
D^A_{ij}=\left|\mu^A_i(S_{ij})-\mu^A_j(S_{ij})\right|.
\]
This produces a human representational dissimilarity matrix (RDM) $D^H$ and an agent RDM $D^A$. RSA is then computed as Spearman correlation between upper-triangular entries of $D^H$ and the permuted $D^A$.

\begin{figure}[!htbp]
    \centering 
    \includegraphics[width=0.5\textwidth]{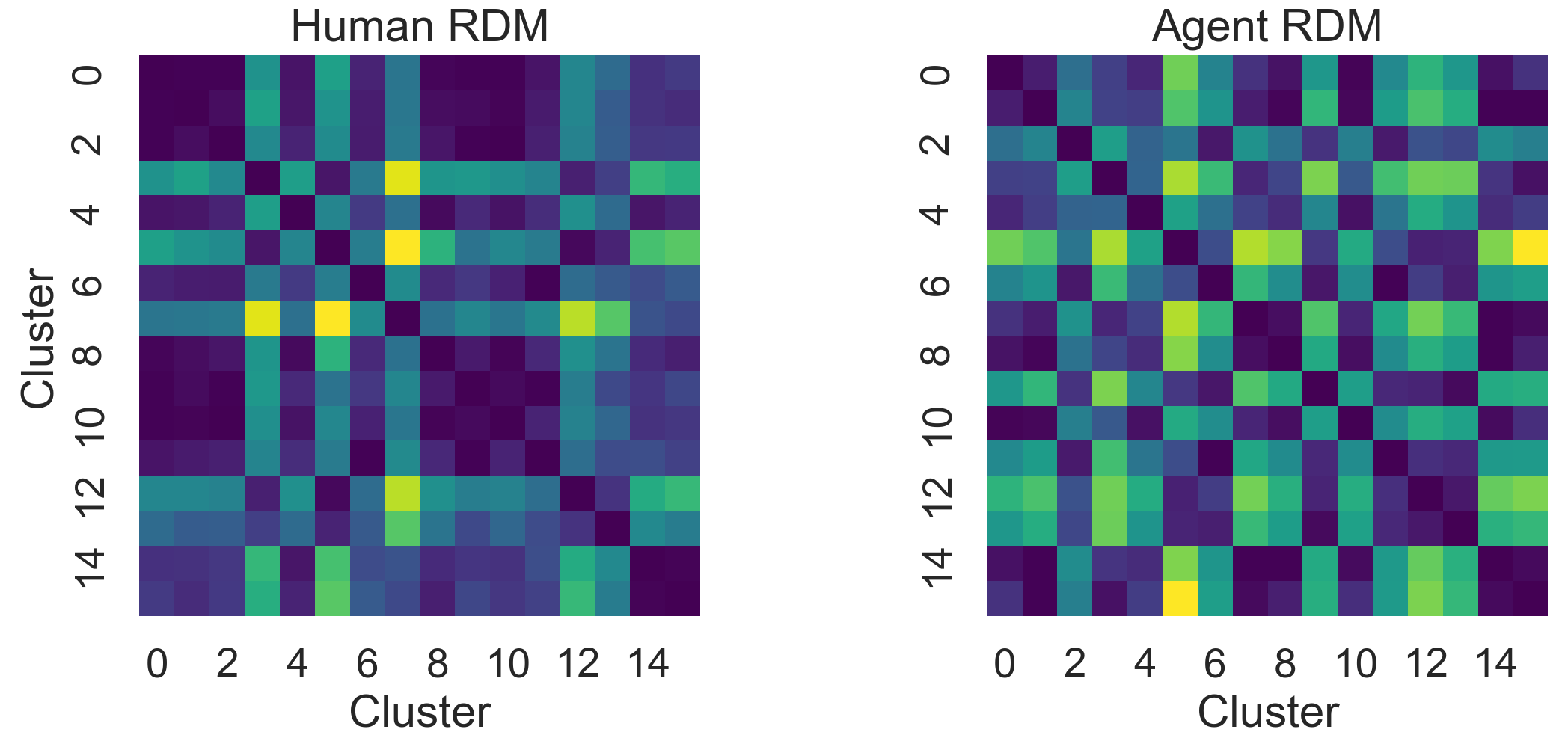}
    \caption{Representation dissimilarity matrices for humans and agents.}
    \label{fig:rdms}
\end{figure}

Significance is assessed by a label-permutation test ($2{,}000$ permutations) on the agent RDM. We obtain
\[
r_s = 0.459,\quad p=0.0015,
\]
(Figure~\ref{fig:alignment} and~\ref{fig:rdms}), indicating statistically significant structural alignment between agent and human cluster-level rating structures. 

\begin{figure*}[!htbp]
    \centering
    \includegraphics[width=\textwidth]{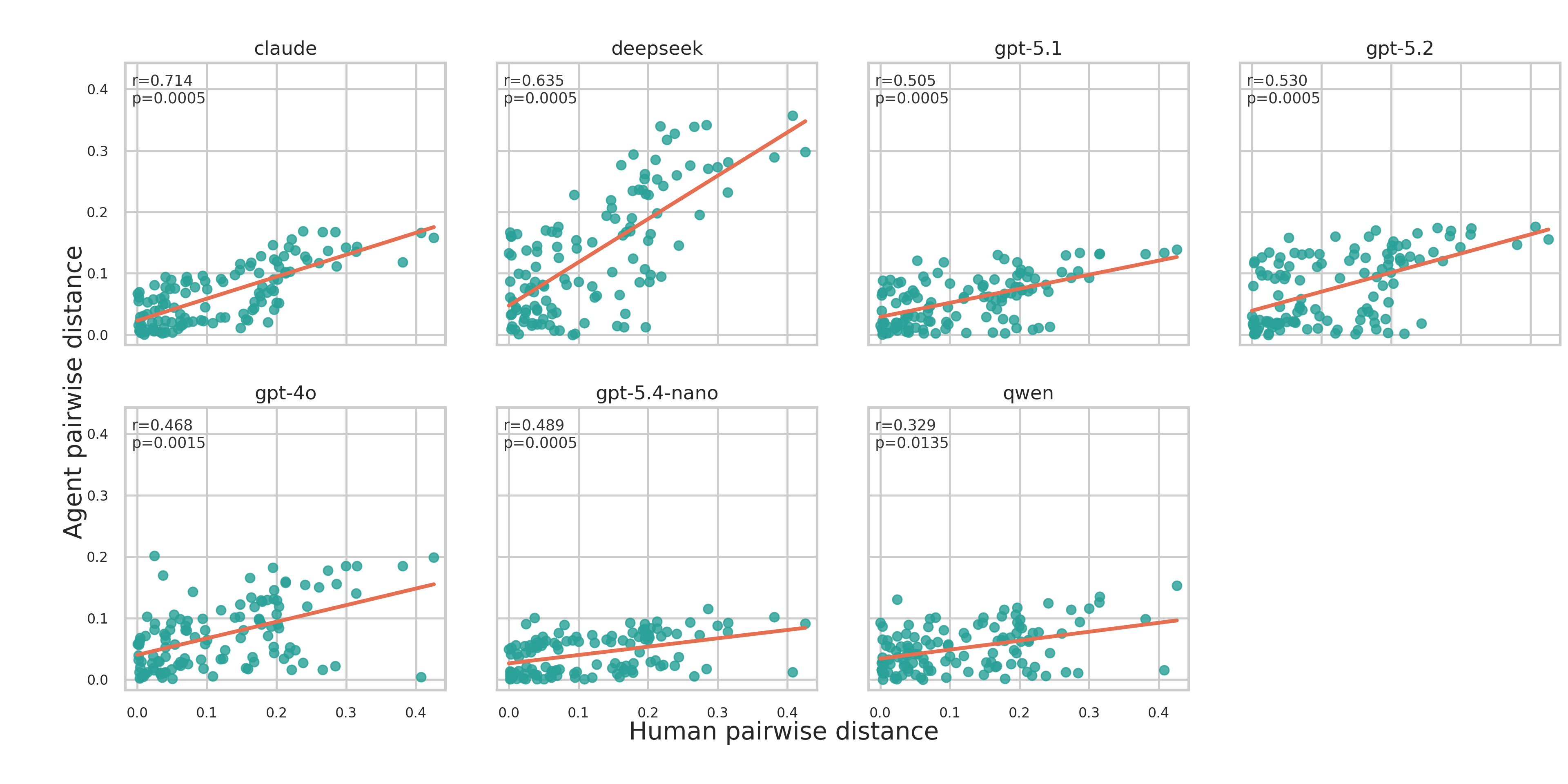}
    \caption{The representational alignment between models and humans across different models, where points corresponded to each number in the corresponding upper triangular indices of the RDM matrices.}
    \label{fig:representational_gen}
\end{figure*}

We further examined the generalizability of these alignments between humans and agents. Using the same methods, we found significant representational similarities among different models and humans, as in Figure~\ref{fig:representational_gen}. Specifically, claude has the highest representational alignment correlation ($r = 0.714$, $p = .0005$), while qwen has the lowest ($r = 0.329$, $p = .0135$). However, all these scores are significant, indicating strong structural alignments between humans and agents.

\begin{figure*}[!htbp]
    \centering 
    \includegraphics[width=\textwidth]{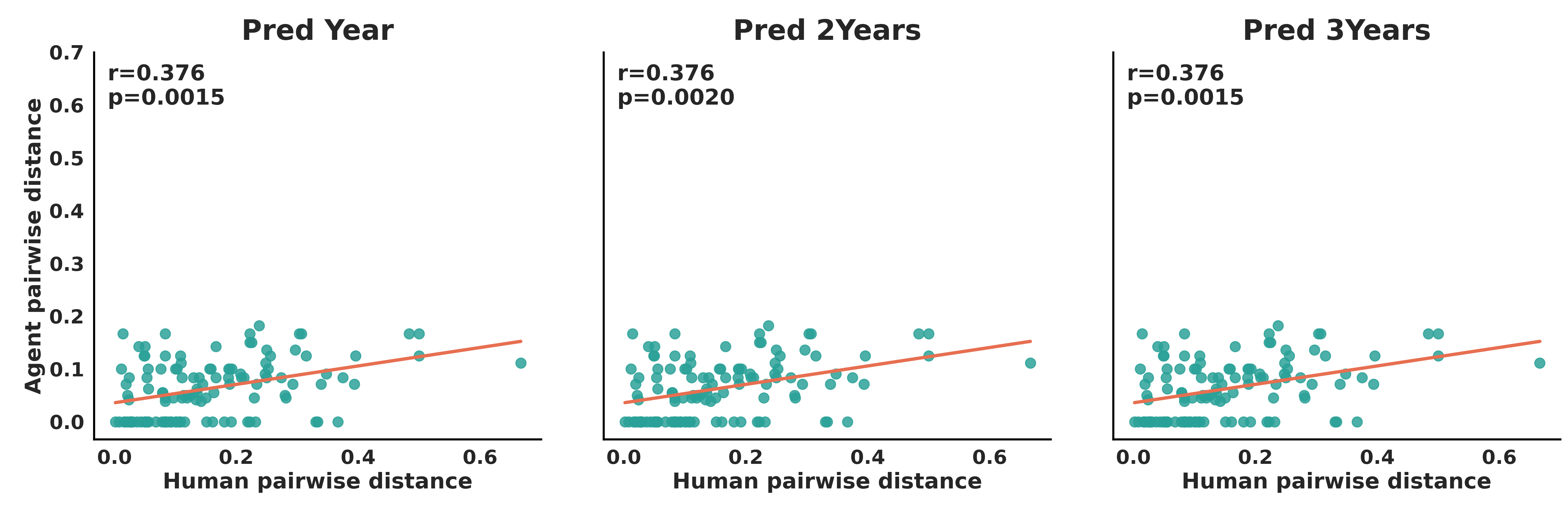}
    \caption{The representational alignment between models and humans across different temporal predictive settings, where points corresponded to each number in the corresponding upper triangular indices of the RDM matrices. }
    \label{fig:representational_temporal}
\end{figure*}

This trend also generalizes to the prediction on future years (e.g., future 1 year to future 3 years). Using still the same methods, we found significant representational similarities among different models and humans, as in Figure~\ref{fig:representational_temporal}. Specifically, these settings feature similar representational alignment correlation and similar significant results (pred 1 year: $r = 0.376$, $p = .0015$; pred 2 years: $r = 0.376$, $p = .002$; pred 3 years: $r = 0.376$, $p = .0015$). This shows the robustness of our persona simulation method.

\subsection{Balanced-set Results}\label{app:balanced_set}

Because the original evaluation set is highly imbalanced, we additionally evaluate MultiCom, two ablated variants, and a single-agent baseline on a sampled balanced 1,998-note set, where the ratio of \texttt{H}/\texttt{NH}/\texttt{NMR} is 1:1:1.
Table~\ref{tab:appendix_balanced_1998_ablation} reports performance by misinformation category.
MultiCom consistently outperforms the single-agent baseline and both ablations, achieving 68.0\% accuracy, 68.0\% balanced accuracy, and 67.3\% Macro-F1 on average.
\begin{table*}[htbp]
    \centering
    \resizebox{\textwidth}{!}{%
    \begin{tabular}{ll cccc cccc}
        \toprule
        \textbf{Method} & \textbf{Metric}
        & \begin{tabular}[c]{@{}c@{}}Manipulated\\ media\end{tabular}
        & \begin{tabular}[c]{@{}c@{}}Factual\\ error\end{tabular}
        & \begin{tabular}[c]{@{}c@{}}Outdated\\ information\end{tabular}
        & \begin{tabular}[c]{@{}c@{}}Missing\\ important context\end{tabular}
        & \begin{tabular}[c]{@{}c@{}}Unverified\\ claim as fact\end{tabular}
        & Satire
        & \begin{tabular}[c]{@{}c@{}}Not\\ misleading\end{tabular}
        & Avg. \\
        \midrule
        
        \multirow{3}{*}{Single Agent}
        & Acc. & 49.2\%$_{3.6}$ & 47.5\%$_{1.7}$ & 53.2\%$_{2.8}$ & 46.8\%$_{1.7}$ & 48.9\%$_{2.3}$ & 54.4\%$_{4.3}$ & 54.9\%$_{2.5}$ & 49.3\%$_{1.1}$ \\
        & Bal. Acc. & 40.0\% & 44.7\% & 50.3\% & 45.6\% & 47.0\% & 47.7\% & 48.4\% & 49.2\% \\
        & Mac. F1 & 38.0\% & 42.2\% & 48.8\% & 43.7\% & 44.0\% & 47.1\% & 30.5\% & 46.2\% \\
        \cmidrule(lr){2-10}
        \multirow{3}{*}{MultiCom w/o Cluster}
        & Acc. & 59.7\%$_{3.5}$ & 59.1\%$_{1.6}$ & 59.4\%$_{2.8}$ & 60.1\%$_{1.6}$ & 62.4\%$_{2.2}$ & 65.0\%$_{4.1}$ & 62.8\%$_{2.4}$ & 60.3\%$_{1.1}$ \\
        & Bal. Acc. & 54.5\% & 57.4\% & 58.4\% & 59.6\% & 61.7\% & 62.7\% & 58.1\% & 60.3\% \\
        & Mac. F1 & 53.9\% & 57.3\% & 58.3\% & 59.1\% & 61.2\% & 62.4\% & 38.5\% & 59.5\% \\

        \cmidrule(lr){2-10}
        \multirow{3}{*}{MultiCom w/o MultiDim}
        & Acc. & 48.7\%$_{3.6}$ & 46.2\%$_{1.7}$ & 46.5\%$_{2.8}$ & 47.2\%$_{1.6}$ & 47.1\%$_{2.2}$ & 54.0\%$_{4.3}$ & 57.5\%$_{2.4}$ & 50.4\%$_{1.1}$ \\
        & Bal. Acc. & 44.6\% & 45.9\% & 46.2\% & 48.3\% & 49.0\% & 47.4\% & 49.7\% & 50.4\% \\
        & Mac. F1 & 40.5\% & 42.1\% & 42.5\% & 44.3\% & 43.9\% & 46.4\% & 26.8\% & 46.1\% \\

        \cmidrule(lr){2-10}
        \multirow{3}{*}{\textbf{MultiCom}}
        & Acc. & 66.5\%$_{3.4}$ & 66.1\%$_{1.6}$ & 68.2\%$_{2.6}$ & 66.2\%$_{1.6}$ & 68.8\%$_{2.1}$ & 72.3\%$_{3.8}$ & 72.9\%$_{2.2}$ & 68.0\%$_{1.0}$ \\
        & Bal. Acc. & 57.5\% & 64.0\% & 66.8\% & 65.3\% & 67.8\% & 67.5\% & 69.5\% & 68.0\% \\
        & Mac. F1 & 57.9\% & 64.3\% & 67.0\% & 65.2\% & 67.8\% & 68.8\% & 46.6\% & 67.3\% \\
        \bottomrule
    \end{tabular}
    }%
    \caption{Ablation and single-agent baseline results on the balanced set. Subscripts denote binomial standard errors for accuracy.}
    \label{tab:appendix_balanced_1998_ablation}
\end{table*}

We further evaluate temporal transfer on the balanced 1,998-note set.
For \textit{Pred Year}, models are trained on notes from year \(t\) and evaluated on notes from year \(t+1\).
For \textit{Pred 2Years}, models are trained on two consecutive years and evaluated on the following two years.
For \textit{Pred 3Years}, models are trained on three consecutive years and evaluated on the following three years.
Table~\ref{tab:appendix_balanced_temporal} reports temporal-transfer accuracy by misinformation category.

\begin{table*}[htbp]
    \centering
    \resizebox{\textwidth}{!}{%
    \begin{tabular}{l cccccccc}
        \toprule
        \textbf{Setting}
        & \begin{tabular}[c]{@{}c@{}}Manipulated\\ media\end{tabular}
        & \begin{tabular}[c]{@{}c@{}}Factual\\ error\end{tabular}
        & \begin{tabular}[c]{@{}c@{}}Outdated\\ information\end{tabular}
        & \begin{tabular}[c]{@{}c@{}}Missing\\ important context\end{tabular}
        & \begin{tabular}[c]{@{}c@{}}Unverified\\ claim as fact\end{tabular}
        & Satire
        & \begin{tabular}[c]{@{}c@{}}Not\\ misleading\end{tabular}
        & Avg. \\
        \midrule
        Pred Year & 45.0\%$_{3.6}$ & 46.4\%$_{1.8}$ & 49.5\%$_{5.0}$ & 50.3\%$_{2.7}$ & 51.7\%$_{6.5}$ & 48.1\%$_{9.6}$ & 59.3\%$_{2.4}$ & 49.7\%$_{1.1}$ \\
        Pred 2Years & 55.7\%$_{2.8}$ & 44.7\%$_{1.4}$ & 45.5\%$_{3.7}$ & 46.7\%$_{2.1}$ & 51.5\%$_{5.0}$ & 39.1\%$_{7.2}$ & 52.4\%$_{1.9}$ & 47.7\%$_{0.9}$ \\
        Pred 3Years & 54.3\%$_{3.9}$ & 48.3\%$_{1.9}$ & 54.7\%$_{5.4}$ & 49.6\%$_{3.0}$ & 55.3\%$_{7.3}$ & 38.1\%$_{10.6}$ & 58.6\%$_{2.6}$ & 51.4\%$_{1.2}$ \\
        \bottomrule
    \end{tabular}
    }%
    \caption{Temporal transfer results on the balanced 1,998-note set. Subscripts denote binomial standard errors for accuracy.}
    \label{tab:appendix_balanced_temporal}
\end{table*}

\section{Methodology Details}
\subsection{Aggregation Feature Views and OOF Predictors}
\label{sec:appendix_aggregation_details}

Table~\ref{tab:appendix_aggregation_features} summarizes the feature views used by the MultiCom aggregator. 
All learned aggregation components are trained in an out-of-fold manner: for each evaluation note, intermediate predictions used by the final ensemble are produced by models that were not trained on that note.

\begin{table*}[htbp]
\centering
\small
\begin{tabularx}{\textwidth}{p{0.23\textwidth} X}
\toprule
Feature view & Included signals \\
\midrule
Raw vote distribution 
& Counts and shares of Helpful; Somewhat Helpful/NMR; Not Helpful; vote entropy; vote margin; resolved-vote mass; NMR gap. \\

Confidence and diagnostic signals 
& Mean confidence; standard deviation of confidence; mean reader-understanding change; confidence-weighted vote shares; label-conditional diagnostic means. \\

Helpfulness reason signals 
& helpfulClear; helpfulGoodSources; helpfulAddressesClaim; helpfulImportantContext; helpfulUnbiasedLanguage. \\

Not-helpfulness reason signals 
& notHelpfulIncorrect; notHelpfulSourcesMissingOrUnreliable; notHelpfulMissingKeyPoints; notHelpfulHardToUnderstand; notHelpfulArgumentativeOrBiased; notHelpfulIrrelevantSources; notHelpfulOpinionSpeculation; notHelpfulNoteNotNeeded. \\

Persona and cluster disagreement 
& Per-agent label indicators; agent-level diagnostic scores; persona-level vote counts; cluster-level vote counts. \\

Note metadata 
& Year; note classification; primary topic; media-note flag; collaborative-note flag; misleading-category flags; non-misleading-category flags; topic count. \\

OOF probability features 
& Out-of-fold class probabilities from summary-feature logistic-regression predictors for NH, NMR, and H; out-of-fold class probabilities from full-agent-feature logistic-regression predictors for NH, NMR, and H. \\
\bottomrule
\end{tabularx}
\caption{Feature views used in the calibrated aggregation process.}
\label{tab:appendix_aggregation_features}
\end{table*}

The final hard ensemble combines five complementary OOF predictors: \texttt{oof\_ensemble\_weighted}, \texttt{oof\_ensemble\_gated}, \texttt{xstyle\_rescue\_gate}, \texttt{blend}, and \texttt{full\_meta}. 
Their weights are empirically optimized as 1.0, 0.75, 0.75, 2.0, and 1.0, respectively. 
We additionally apply a promotion rule for conservative \texttt{NMR} predictions: when the rationale/metadata blend and structured-summary predictor agree on the same resolved label, and vote-level diagnostic thresholds are satisfied, the prediction is promoted from \texttt{NMR} to that resolved label.
For the temporal generalization experiment presented in Table~\ref{tab:generalizability}, we construct year-based rolling train-test splits based on the creation year of each note. 
This setting differs from the primary 5-fold out-of-fold evaluation approach because under this configuration, test notes are always temporally later than training notes. 
For \textit{Pred Year}, candidate windows are defined as \(t \rightarrow t+1\), where the model is trained and calibrated on notes from year \(t\) and evaluated on notes from year \(t+1\). 
For \textit{Pred 2Years}, candidate windows are \((2021,2022)\rightarrow(2023,2024)\), \((2022,2023)\rightarrow(2024,2025)\), and \((2023,2024)\rightarrow(2025,2026)\). 
For \textit{Pred 3Years}, candidate windows are \((2021,2022,2023)\rightarrow(2024,2025,2026)\). 

Within each training window, all learned aggregation components are fitted solely on notes from the training years. 
The component predictors are calibrated through out-of-fold splits internal to the training window, and the aforementioned fixed hard ensemble rules are applied to the test windows corresponding to future years subsequently. 
Notes from the future test years are strictly excluded from the fitting of component models, the selection of thresholds, and the application of boosting rules.
If the number of samples from all three status categories contained within the training years is insufficient to support internal calibration, the corresponding candidate window is skipped.
Finally, prior to computing the category-level accuracies reported in Table~\ref{tab:generalizability}, predictions from all feasible temporal windows under the same setting are pooled.

\subsection{Auxiliary OOF Predictors and Promotion Rule}

The final ensemble uses several out-of-fold predictors, each trained without access to the held-out notes in its fold. 
The summary-metadata predictor takes note-level summary features as input, including vote shares, vote entropy, vote margin, confidence statistics, agreement rates, helpfulness-reason rates, not-helpfulness-reason rates, and metadata-derived OOF probabilities. 
The structured-summary predictor uses the same summary-metadata view, augmented with structured diagnostic features aggregated from agent outputs, including rationale length, reader-understanding scores, persona-level vote counts, and cluster-level vote counts. 
The full-metadata predictor further includes per-agent label indicators and per-agent diagnostic scores. 
The rationale-text predictor uses concatenated agent rationales with lightweight text features. 
The blend predictor is an OOF probability blend of the summary-metadata, structured-summary, full-metadata, and rationale-text predictors.

For the final hard ensemble, we combine five OOF anchors: \texttt{oof\_ensemble\_weighted}, \texttt{oof\_ensemble\_gated}, \texttt{xstyle\_rescue\_gate}, \texttt{blend}, and \texttt{full\_meta}, with weights 1.0, 0.75, 0.75, 2.0, and 1.0, respectively. 
After this ensemble prediction, we apply a conservative promotion rule only when the initial prediction is \texttt{NMR}. 
Specifically, if the blend predictor and the structured-summary predictor agree on the same resolved label, and the label is not \texttt{NMR}, we promote the final prediction to that label only when the NMR vote share is at least 0.56 and the mean reader-understanding score is at most 29.79. 
These thresholds are applied to vote-level diagnostics and are used to recover high-confidence resolved cases while avoiding promotion based on a single auxiliary signal.

\subsection{Fine-tuned Baseline Implementation Details}
\label{sec:appendix_finetune_details}

We implement the fine-tuned baseline as a direct three-class classifier over \texttt{NH}, \texttt{NMR}, and \texttt{H}. 
The backbone is Mistral-7B-Instruct-v0.3, fine-tuned with LoRA~\cite{hu2022lora}. 
The fine-tuning set contains 106,611 ComRate examples with available post text, note text, and official status labels, including 4,211 \texttt{NH}, 94,538 \texttt{NMR}, and 7,862 \texttt{H} examples. 
Each input concatenates the post, the community note, and a three-way classification instruction.

Similar to Xing et al.~\cite{xing2026communitynotes}, we pool the final-token hidden representation from the backbone and feed it into a linear classification head. 
Training uses weighted cross-entropy with square-root inverse class weights and a weighted random sampler. 
We use five-fold cross-validation for determining best checkpoints. In each fold, the held-out fold is used for testing.

We empirically determine the hyperparameters as follows: maximum sequence length 512, batch size 2, gradient accumulation 4, learning rate \(2\times10^{-4}\), weight decay 0.01, warmup ratio 0.1, 3 epochs, LoRA rank 16, LoRA alpha 32, and LoRA dropout 0.05. 

\subsection{Generalizabilty Setting Justifications}\label{app:generalizability}

We evaluated three aspects of generalizability, across different models, across different temporal aspects, and across different agent numbers. 

For different models, we tested models with different brands regions. These potentially resulted in different processing capabilities in different language environments. Besides, different brands' models may have different reply tendencies, or even different personalities, experiments with different models could test whether the persona simulation framework is generalizable. Therefore, we selected GPT-4o, GPT-5.1, Claude Opus 4.6, Qwen3-Max and DeepSeek-v3.2. Notably, this experiment is not meant to be exhaustive, and we note that there are varied open-sourced models, which we could expand to test in the future.

Besides, for generalizability along the temporal aspect, as the dataset starts from 2021 to 2026, we decided that the longest temporal window is to use the past three years data (2021-2023) to predict the next three years data (2024-2026). Therefore, we dedicated the longest prediction time window is \textit{Pred 3Years}, and thereby designed three settings: \textit{Pred Year}, \textit{Pred 2Years}, \textit{Pred 3Years}. For these settings, we used all past data for training or persona calibration, and predicted on all the following years' (e.g., 1 year, 2 years or 3 years) notes.

Finally, we evaluated scaling behavior by configuring the system with 16, 32, and 48 agents. We selected this range because scaling from 16 to 48 agents already revealed distinct performance trends: a consistent increase in \texttt{H}/\texttt{NH} precision and overall accuracy, alongside a decline in balanced accuracy. Crucially, even the 16-agent configuration substantially outperformed the 1-agent baseline across all three metrics, establishing 16 agents as a highly effective baseline setting. Conversely, scaling beyond 48 agents (e.g., to 64 or more) incurs prohibitive computational and financial costs for single-note prediction. We therefore restricted our evaluation to these three configurations to balance performance insights with practical efficiency.

\subsection{Fine-grained Label Predictions}\label{app:label}

For fine-grained label prediction, we use the structured outputs already generated by MultiCom, including vote distributions, confidence, helpfulness and quality signals, failure mode signals, and disagreement patterns. Subsequently, we train an out-of-fold multi-label classifier to predict the official reason labels for each note. This setup is designed to assess whether the diagnostic dimensions extracted from simulated raters align with the explanatory labels provided by human annotators. Although MultiCom is capable of generating a broader spectrum of diagnostic signals regarding justification levels during the agent simulation process, four official prediction labels were excluded from the primary evaluation because their low frequency in the 2,000-note evaluation set, making their precision, recall, and F1 scores unstable.

\end{document}